\documentclass[%
 aip,
 amsmath,amssymb,
 reprint,%
]{revtex4-1}

\usepackage{graphicx}
\usepackage{dcolumn}
\usepackage{bm}

\usepackage[utf8]{inputenc}
\usepackage[T1]{fontenc}
\usepackage{mathptmx}

\begin{document}

\preprint{AIP/123-QED}

\title{Stabilization of steady state in Multiplex heterogeneous networks of neuron-like models with bistability between silent state and bursting attractor}

\author{N. Stankevich}
\email{stankevichnv@mail.ru.}
 \affiliation{HSE University, 25/12 Bolshay Pecherskaya str., Nizhny Novgorod 603155, Russia}


\date{\today}

\begin{abstract}
The dynamics of a multiplex heterogeneous network of oscillators is studied. Two types of similar models based on the Hodgkin-Huxley formalism are used as the basic elements of the networks. The first type model demonstrates bursting oscillations. The second model demonstrates bistability between bursting attractor and stable steady state. Basin of attraction of the stable equilibrium in the model is very small. Bistabilty is a result taking into account an additional ion channel, which has a non-monotonic characteristic and can be interpreted as a channel with a communication defect. Suggested multiplex networks assumed more active communication between models with a defect as a result in such networks it is enough to have one element with a communication defect in the subnetworks in order to stabilize the state of equilibrium in the entire network.
\end{abstract}

\maketitle


\section{\label{sec:level1}Introduction}

Networks of interacting oscillators are one of the most important research objects of the dynamics of complex systems in various fields of science \cite{bornholdt2001handbook,strogatz2001exploring,dorogovtsev2002evolution,newman2003structure,kirst2016dynamic,zhang2020topological}. One of the most interesting areas is the study of the interaction of models of neurons described by the Hodgkin-Huxley formalism \cite{hodgkin1952currents}, since it is directly related to the study of the interaction of biological cells, and is also important for the development of artificial intelligence and machine learning \cite{izhikevich2007dynamical, shen2021multistability}.

Typical behavior corresponding to the normal mode of cell functioning described by the Hodgkin-Huxley formalism, such as neurons, pancreatic beta cells, cardiomyocytes, and others, is a dynamic mode corresponding to the bursting attractor. Bistability can be observed in such systems \cite{shen2021multistability}. Moreover, different types of attractors can coexist, various types of multistability are discussed in the papers  \cite{malashchenko2011six,stankevich2018stochastic}. One of the most interesting options is the multistability between a bursting attractor and a stable equilibrium state. In \cite{stankevich2017coexistence}, a modification of the well-known Sherman model \cite{sherman1988emergence, sherman1992rhythmogenic} is proposed, in which, in addition to a typical bursting attractor, an equilibrium state is stabilized. The modification of the model consists in taking into account an additional ion channel with a non-monotonic characteristic, which locally changes the nullcline of the model's fast manifold, i.e. the model retains all of its basic properties. A new ion channel can be interpreted as a defect in cell communication, since the probability of its opening is lower than that of a conventional channel. The presence of similar models makes it possible to model heterogeneous networks, some of the elements of which have a communication defect, and some do not \cite{stankevich2020cooperative}. From the point of view of dynamical systems, such a situation will correspond to the fact that some of the elements of the network will demonstrate only a burst attractor, while the equilibrium state will be unstable. And some of them will also mainly demonstrate bursting oscillations, while a stable state of equilibrium with a small basin of attraction will coexist with them. The study of the dynamics of a heterogeneous network of interacting models showed \cite{stankevich2020cooperative} that with a ratio of elements with and without pathology in a ratio of 1:1, the state of equilibrium will be unstable and pathology will not appear in the system. With an increase in the number of pathological elements (with bistability), the equilibrium state can be stabilized, however, the stabilization threshold for the coupling strength parameter increases with a decrease in the ratio of normal elements to elements with pathology, which indicates the presence of resistive properties in such systems and preservation of the normal mode of cell functioning.

In this work, we will consider a multiplex heterogeneous network. We will assume that abnormal elements (with bistability) may behave more actively than normal ones. In the context of such an assumption, the ratio of defective and non-defective elements can change to manifest pathological behavior.

The work is structured as follows. In Section 2, we describe a model with and without a communication defect and demonstrate typical dynamic modes. In Section 3, we present the results of numerical modeling of a multiplex heterogeneous network with the same number of elements in a subnetwork, describe the assumptions of a more active behavior of cells with pathology, and present the results of modeling the dynamics of rather small networks of this type. In Section 4, we present the results of a study of networks with different numbers of elements in subnetwork.


\section{\label{sec:2} Object of investigation}

To depict the bursting dynamics as observed under physiological conditions, as well as to simultaneously model silent dynamics caused upon ion channel disregulation, we use a model based on the Hodgkin-Huxley formalism with modifications proposed in \cite{stankevich2017coexistence}:

\begin{equation}
\begin{array}{l l l}
\tau \dot{V} = -I_{C_a}(V) - I_K(V,n) - k I_{K2}(V)-I_S(V,S) \\
\tau \dot{n} = \sigma (n_{\infty}(V) - n) \\
\tau_S \dot{S} = S_{\infty}(V) - S
\end{array}
\label{eq:main_model}
\end{equation}

where $V$ represents the membrane potential of the cell, the functions $I_{C_a}(V)$, $I_K(V,n)$, $I_S(V, S)$ define the three intrinsic currents, fast calcium, potassium and slow potassium respectively, such that:

\begin{equation}
I_{C_a}(V) = g_{C_a}m_{\infty}(V)(V-V_{C_a})
\end{equation}
\begin{equation}
I_{K}(V,n) = g_{K}n(V-V_{K})
\label{eq:normal_current}
\end{equation}
\begin{equation}
I_{S}(V,n) = g_{S}S(V-V_{K}).
\end{equation}

The gating variables for $n$ and $S$ are the opening probabilities of the fast and slow potassium currents:

\begin{equation}
\omega_{\infty}(V) = [1+exp\frac{V_{\omega}-V}{\Theta_{\omega}}]^{-1}, \omega=m,n,S
\label{eq:probab}
\end{equation}

\begin{figure}[htbp!]
\centering
\includegraphics[width=0.8\columnwidth]{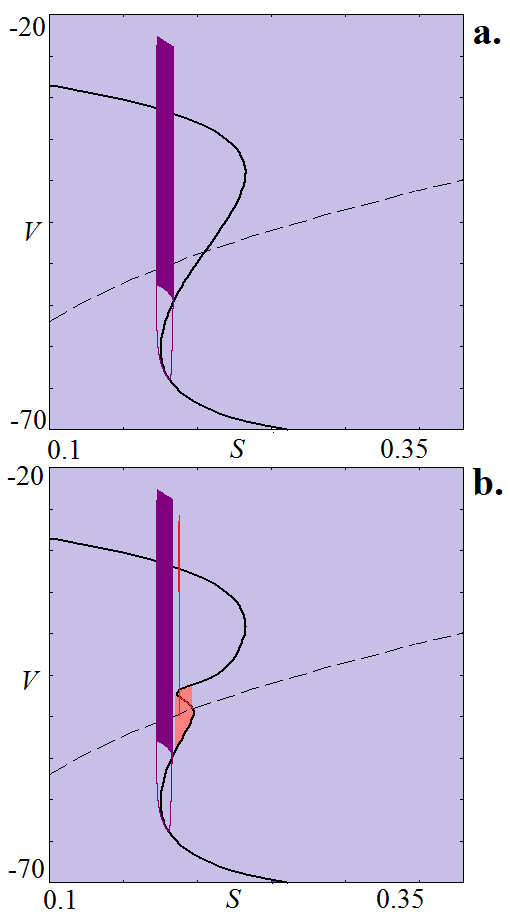}
\caption{Dynamical structure of the model (\ref{eq:main_model}) in the regime bursting (\textbf{a.}) and bistability (\textbf{b.}). Fast and slow manifolds (black solid and dashed lines, respectively), attractors, and their basins of attraction are shown for k = 0 (\textbf{a.}) and k = 1 (\textbf{b.}). Parameters: $\tau$ = 0.02, $\tau_S$ = 35, $\sigma$ = 0.93, $g_{Ca} = 3.6$, $g_K = 10.0$, $g_{K2} = 0.2$, $g_S$ = 4.0, $V_Ca$ = 25.0, $V_K$ = -75.0, $V_m$ = -20.0, $\theta_m$ = 12.0, $V_n$ = -16.0, $\theta_n$ = 5.6, $V_S$ = -35, $\theta_S$ = 10.0, $V_P$ = -47.0, $\theta_p$ = 1.0.}
\label{fig:1}
\end{figure}

The function $I_{K2}(V)$ on the other hand defines an additional voltage-dependent potassium current. It has been previously demonstrated that a single oscillator, in absence of $I_{K2}$ (when $k=0$), is characterized with a bursting attractor that is born via a Hopf bifurcation for $V_S=-44.7mV$. At this parameter value the equilibrium point looses its stability. Although the bursting attractor is born in the vicinity of the equilibrium point, this point moves away from the bursting attractor for increase in $V_S$. Even more, the two dimensional projection of the phase portrait together with the fast and slow manifolds when $V_S=-35mV$ (Fig.~\ref{fig:1}a) demonstrates that the periodic trajectories do not intersect the neighborhood of the steady state and the bursting attractor terminates in a homoclininc bifurcation as the trajectory hits the slow manifold \cite{izhikevich2007dynamical}.

In the presence of $I_{K2}(V)$ ($k \neq 0$) that varies strongly with the membrane potential in the vicinity of the equilibrium point however, the unstable node is stabilized without affecting the global flow of the model (Fig.~\ref{fig:1}b)\cite{stankevich2017coexistence}. $I_{K2}(V)$ is given in the form:

\begin{equation}
I_{K2}(V) = g_{K2}p_{\infty}(V)(V-V_K)
\label{eq:new}
\end{equation}

where

\begin{equation}
p_{\infty}(V)=[exp\frac{V-V_p}{\Theta_p}+exp\frac{V_p-V}{\Theta_p}]^{-1}
\label{eq:probab1}
\end{equation}

represents the opening probability of this channel. In contrast to the other potassium channel (Eqs. \ref{eq:normal_current}, \ref{eq:probab}) which opens with probability $n_{\infty}=1$ when the membrane voltage reaches a threshold value, the opening probability of the modified channel will be equal only to 0.5. From physiological point of view, this can be interpreted as an ion channel disfunction, as for instance blocking of the potassium channel or its inactivation [ref], and thereby a stable silent state emerges. Between the stable node and the bursting attractor there is a rejecting current which enables the system to remain in the stable steady state when starting from initial conditions in its vicinity. Generally, the modified model ($k \neq 0$) depicts bistability between physiological, bursting dynamics and a pathological silent state dynamics.

\section{Symmetric multiplex network}
\label{sec:3}
To investigate the global dynamics of a network cells of model (\ref{eq:main_model}),
we introduced electrical coupling by adding the following gap- junctional coupling term to the equation for $V$ in Eq. (\ref{eq:main_model}) that describes bidirectional transport of ions between the cells

\begin{equation}
I_C(V^{(i)})=\sum_{j \in \Gamma_i}g_{Ce}(V^{(i)}-V^{(j)})
\end{equation}
with $g_{Ce}$ being the coupling conductance (coupling strength). In this way, only electrical coupling of the cells is considered. The sum is taken over the whole population, assuming global coupling. As already noted, if $k = 0$ and in the absence of coupling, the isolated oscillators display a single stable state-bursting dynamics (Fig.~\ref{fig:1}a), if $k = 1$ and in the absence of coupling, the isolated oscillators display a bistability between bursting dynamics and stable steady state (Fig.~\ref{fig:1}b).

In the present work we will consider multiplex heterogeneous network. For it we will introduce two types of coupling, one is intra-subnetwork and the other is between subnetwork. Moreover, our assumption that elements with pathology behave more actively is that communication between subnetworks is carried out only through elements with a defect. The model of such a network can be written as follows:
\begin{equation}
\begin{array}{l l}
I_C(V^{(i)})=\sum_{j \in \Gamma1_i}g_{Ce}^{In}(V^{(i)}-V^{(j)})+\\
\sum_{j \in \Gamma2_i}g_{Ce}^{In}(V^{(i)}-V^{(j)})+\sum_{j \in \Gamma3_i}g_{Ce}^{Out}(V^{(i)}-V^{(j)})
\label{eq:multiplex_coupl}
\end{array}
\end{equation}
with $g_{Ce}^{In}$ being the coupling strength inside subnetworks $\Gamma 1$ and $\Gamma 2$ and $g_{Ce}^{Out}$ being the coupling strength between subnetworks implemented with elements with defect from another subnetwork $\Gamma 3$.

\subsection{Minimal network: $N=4$}

The minimal network of interacting oscillators consists of two interacting elements in \cite{stankevich2020cooperative}, the main types of behavior of two coupled oscillators (\ref{eq:main_model}) were considered and described. Moreover, all variants of interacting oscillators were considered depending on the coefficient $k$. Stable equilibrium is possible only in the case of interaction of two oscillators with bistability ($k_1$ = $k_2$ = 1). If one oscillator has a communication defect ($k_1$ = 1), and the other does not ($k_2$ = 0), then the equilibrium state is always unstable. We will use such a minimal heterogeneous network of different elements as a subnetwork and consider the interaction of two such subnetworks. Thus, the complete network will consist of four elements, two of which have communication defect and two have not. In such a network with a global coupling, it is impossible to stabilize the equilibrium state, it will always be unstable.

\begin{figure}[htbp!]
\centering
\includegraphics[width=0.8\columnwidth]{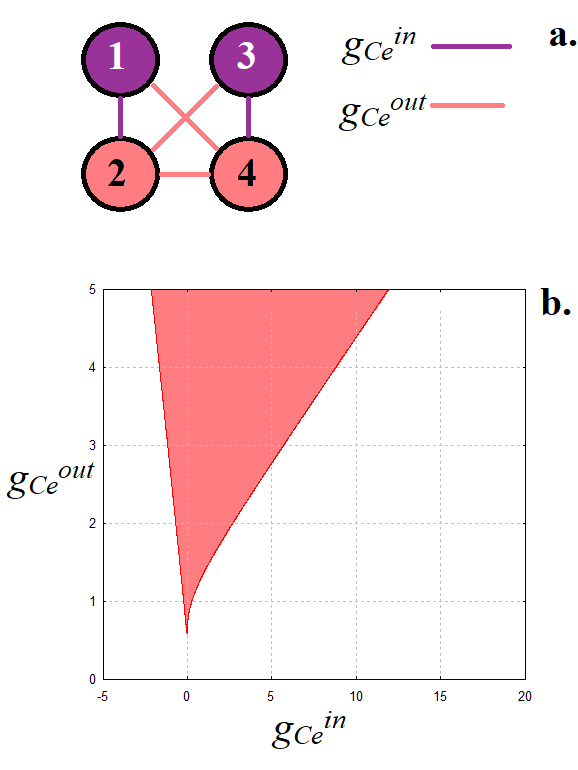}
\caption{Schematic representation of coupled networks (\textbf{a.}) and bifurcation diagram for model (\ref{eq:4cell_mn}) (\textbf{b.}). Parameters: $\tau$ = 0.02, $\tau_S$ = 35, $\sigma$ = 0.93, $g_{Ca} = 3.6$, $g_K = 10.0$, $g_{K2} = 0.2$, $g_S$ = 4.0, $V_Ca$ = 25.0, $V_K$ = -75.0, $V_m$ = -20.0, $\theta_m$ = 12.0, $V_n$ = -16.0, $\theta_n$ = 5.6, $V_S$ = -35, $\theta_S$ = 10.0, $V_P$ = -47.0, $\theta_p$ = 1.0. $k_1=0$, $k_2=1$, $k_3=0$, $k_4=1$. Red notes have communication defect and demonstrate bistability between silent state and bursting oscillations.}
\label{fig:2}
\end{figure}

\begin{figure}[htbp!]
\centering
\includegraphics[width=0.9\columnwidth]{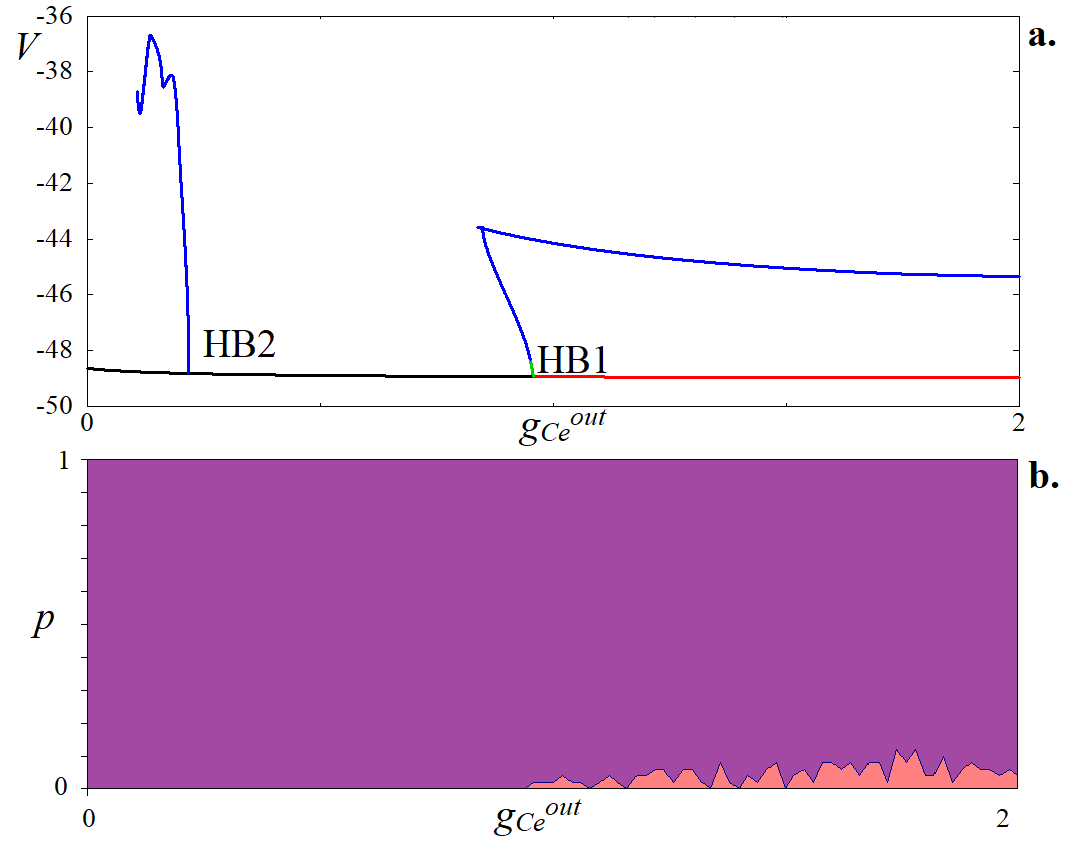}
\caption{One-parametric bifurcation diagram (\textbf{a.}) and probability of coexisting attractors (\textbf{b.}) for model (\ref{eq:4cell_mn}). Parameters: $\tau$ = 0.02, $\tau_S$ = 35, $\sigma$ = 0.93, $g_{Ca} = 3.6$, $g_K = 10.0$, $g_{K2} = 0.2$, $g_S$ = 4.0, $V_Ca$ = 25.0, $V_K$ = -75.0, $V_m$ = -20.0, $\theta_m$ = 12.0, $V_n$ = -16.0, $\theta_n$ = 5.6, $V_S$ = -35, $\theta_S$ = 10.0, $V_P$ = -47.0, $\theta_p$ = 1.0. $k_1=0$, $k_2=1$, $k_3=0$, $k_4=1$, $g_{Ce}^{in} = 0.2.$. Red notes have communication defect and demonstrate bistability between silent state and bursting oscillations.}
\label{fig:3}
\end{figure}

We developed multiplex heterogeneous network, in corresponding assumption (\ref{eq:multiplex_coupl}) we introduced two types of couplings, one is intra-subnetworks and the other is between subnetwork. Moreover, our assumption that elements with pathology behave more actively is that communication between subnets is carried out only through elements with a defect. Figure \ref{fig:2} shows a schematic representation of such a network. The model of such a network can be written as follows:

\begin{equation}
\begin{array}{l r l l l l l l}
\tau \dot{V_1} = -I_{C_a}(V_1) - I_K(V_1,n_1) - k_1 I_{K2}(V_1)-I_S(V_1,S_1)\\ - g_{Ce}^{In}(V_2-V_1) - g_{Ce}^{Out}(V_4-V_1),\\
\tau \dot{V_2} = -I_{C_a}(V_2) - I_K(V_2,n_2) - k_2 I_{K2}(V_2)-I_S(V_2,S_2)\\ - g_{Ce}^{In}(V_1-V_2) - g_{Ce}^{Out}(V_4-V_2),\\
\tau \dot{V_3} = -I_{C_a}(V_3) - I_K(V_3,n_3) - k_3 I_{K2}(V_3)-I_S(V_3,S_3)\\ - g_{Ce}^{In}(V_4-V_3) - g_{Ce}^{Out}(V_2-V_3),\\
\tau \dot{V_4} = -I_{C_a}(V_4) - I_K(V_4,n_4) - k_4 I_{K2}(V_4)-I_S(V_4,S_4)\\ - g_{Ce}^{In}(V_3-V_4) - g_{Ce}^{Out}(V_2-V_4),\\
\end{array}
\label{eq:4cell_mn}
\end{equation}

where $k_1 = k_3 = 0.0$ the first and the third models (one of the cells in subnetworks) have not defect communication and manifest only bursting oscillations, $k_2 = k_4 = 1.0$ the second and the forth models have defect and demonstrate bistability between stable steady state and bursting attractor.

An analysis of the stability of the equilibrium state was carried out using the XPPAUT software package \cite{ermentrout2002simulating} depending on the coupling coefficients. Figure \ref{fig:2}b shows the bifurcation diagram for model (\ref{eq:4cell_mn}); the area where the equilibrium state is stable is marked in pink. The abscissa shows the parameter responsible for the strength of communication within the subnetwork, and the ordinate shows the parameter responsible for the communication between subnets, implemented by elements with pathology. Figure \ref{fig:2}b clearly shows that the area of equilibrium stabilization has a threshold in terms of the parameter of communication between subnets, for $g_{Ce}^{in} = 0$, the least strength of communication between subnets is required to stabilize the equilibrium. With an increase in communication within the subnets, the stabilization threshold increases. Thus, the interaction within the subsystems interferes with stabilization and stronger interaction between the subnets is required for stabilization. At the same time, as we see, stabilization is possible in a large area of the parameter space of communication. Thus, in a heterogeneous globally coupled network only unstable equilibrium will be observed. More active behavior of elements with pathology can contribute to the occurrence of pathological conditions in a network with the same number of elements with and without pathology. If the pathological element is replaced with a normal one in one of the subnets, the equilibrium state will destabilize. Thus, for a model of 4 cells, at least 50\% of elements with pathology are required for it to manifest itself in the entire network.

In Fig.~\ref{fig:3}a one parametric bifurcation diagram is presented. Stable equilibrium lost stability as a result Andronov-Hopf bifurcation at  $g_{Ce}^{out} \approx 0.96$ and a limit cycle is born. With further decreasing in coupling strength between subnets there is one more Andronov-Hopf bifurcation at $g_{Ce}^{out} \approx 0.22$, which gives saddle cycle. This two cycles form bursting attractor. Fig.~\ref{fig:3}b probability characteristic of coexistence attractors shows. For each value of coupling strength $g_{Ce}^{out}$ we take 50 random initial conditions inside the phase space volume covering coexisting attractors, and calculate probability to get stable equilibrium. Probability of stable equilibrium less the 20\%.

\subsection{Bigger network: $N=6$}

\begin{figure*}[htb!]
\centering
\includegraphics[width=0.7\textwidth]{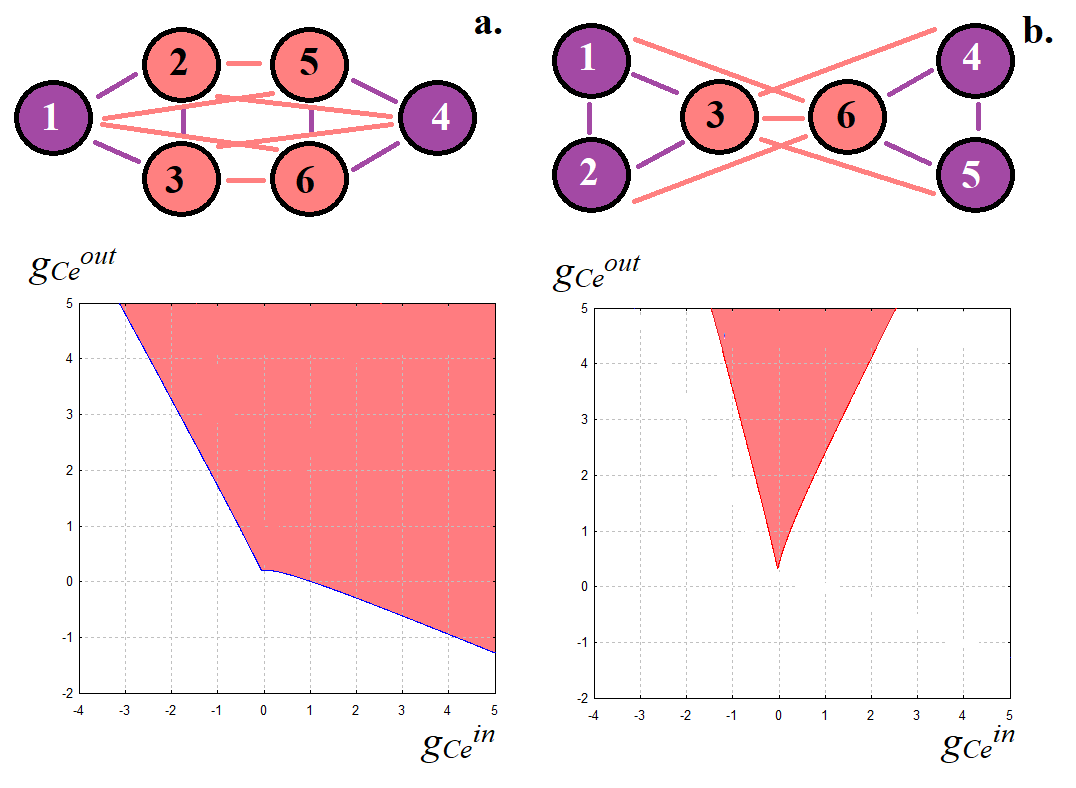}
\caption{Schematic representation of coupled networks and bifurcation diagram for model (\ref{eq:6cell_mn}). Parameters: $\tau$ = 0.02, $\tau_S$ = 35, $\sigma$ = 0.93, $g_{Ca} = 3.6$, $g_K = 10.0$, $g_{K2} = 0.2$, $g_S$ = 4.0, $V_Ca$ = 25.0, $V_K$ = -75.0, $V_m$ = -20.0, $\theta_m$ = 12.0, $V_n$ = -16.0, $\theta_n$ = 5.6, $V_S$ = -35, $\theta_S$ = 10.0, $V_P$ = -47.0, $\theta_p$ = 1.0. $k_1=0$, $k_2=1$, $k_3=0$, $k_4=1$. \textbf{a.} $k_1=0$, $k_2=1$, $k_3=0$, $k_4=1$. \textbf{b.}Red notes have communication defect and demonstrate bistability between silent state and bursting oscillations.}
\label{fig:4}
\end{figure*}

Next, we will increase the size of the subnets by one element. Thus, there will be three elements in the subnet. In such a subnet, we can already implement pathological behavior in the presence of 2 elements with pathology and one normal one. In order to stabilize the state of equilibrium in such a system with global coupling parameter of coupling strength more than 1.028 is required \cite{stankevich2020cooperative}. As in the previous section, we will develop a multiplex heterogeneous network, in which the interaction between subsystems will be mediated only by elements with pathology. Thus, we get a system consisting of 6 nodes, among which 4 elements with pathology and 2 without. In \cite{stankevich2020cooperative}, a study of such a system with a global coupling is presented and it is shown that stabilization will be observed at  $g_{Ce} = 0.514$. We write the multiplex network as follows:

\begin{equation}
\begin{array}{l l l l l l l l l l l l}
\tau \dot{V_1} = -I_{C_a}(V_1) - I_K(V_1,n_1) - k_1 I_{K2}(V_1)-I_S(V_1,S_1) - \\ g_{Ce}^{In}(V_2+V_3-2 V_1) - g_{Ce}^{Out}(k_4 V_4 + k_5 V_5 + k_6 V_6 - (k_4+k_5+k_6)V_1),\\
\tau \dot{V_2} = -I_{C_a}(V_2) - I_K(V_2,n_2) - k_2 I_{K2}(V_2)-I_S(V_2,S_2) - \\ g_{Ce}^{In}(V_1+V_3-2 V_2) - g_{Ce}^{Out}(k_4 V_4 + k_5 V_5 + k_6 V_6 - (k_4+k_5+k_6)V_2),\\
\tau \dot{V_3} = -I_{C_a}(V_3) - I_K(V_3,n_3) - k_3 I_{K2}(V_3)-I_S(V_3,S_3) - \\
g_{Ce}^{In}(V_1+V_2-2 V_3) - g_{Ce}^{Out}(k_4 V_4 + k_5 V_5 + k_6 V_6 - (k_4+k_5+k_6)V_3),\\
\tau \dot{V_4} = -I_{C_a}(V_4) - I_K(V_4,n_4) - k_4 I_{K2}(V_4)-I_S(V_4,S_4) - \\
g_{Ce}^{In}(V_5+V_6-2 V_4) - g_{Ce}^{Out}(k_1 V_1 + k_2 V_2 + k_3 V_3 - (k_1+k_2+k_3)V_4),\\
\tau \dot{V_5} = -I_{C_a}(V_5) - I_K(V_5,n_5) - k_5 I_{K2}(V_5)-I_S(V_4,S_4) - \\
g_{Ce}^{In}(V_4+V_6-2 V_5) - g_{Ce}^{Out}(k_1 V_1 + k_2 V_2 + k_3 V_3 - (k_1+k_2+k_3)V_5),\\
\tau \dot{V_6} = -I_{C_a}(V_6) - I_K(V_6,n_6) - k_6 I_{K2}(V_6)-I_S(V_6,S_6) - \\
g_{Ce}^{In}(V_4+V_5-2 V_6) - g_{Ce}^{Out}(k_1 V_1 + k_2 V_2 + k_3 V_3 - (k_1+k_2+k_3)V_6),\\
\end{array}
\label{eq:6cell_mn}
\end{equation}

Figure \ref{fig:4}a shows a schematic representation of a network for $k_1=0$, $k_2=1$, $k_3=1$, $k_4=0$, $k_5=1$, $k_6=1$, as well as a bifurcation diagram for such a network. As expected, there is no stabilization threshold for such a network even at $g_{Ce}^{out} = 0$, with an increase in the coupling strength within the clusters, the equilibrium state is stabilized at $g_{Ce}^{in} \approx 1$. This threshold is two times larger then for globally coupled similar network.

Now let us change the subnets in such a way that two will be normal, and one will remain with pathology, in model (\ref{eq:6cell_mn}) it corresponds to $k_1=0$, $k_2=0$, $k_3=1$, $k_4=0$, $k_5=0$, $k_6=1$. Figure \ref{fig:4}b shows a schematic representation of interacting elements, as well as a bifurcation diagram. Thus, in this case, despite the fact that in the complete network the number of elements with pathology is two times less than the normal ones, stabilization of the pathological equilibrium state is possible. The shape of the region in the parameter space is similar to that was for 4-cell model (\ref{eq:4cell_mn}) in Fig.~\ref{fig:2}.

\section{\label{sec:4} Asymmetric multiplex network}

\begin{figure*}[htb!]
\centering
\includegraphics[width=0.9\textwidth]{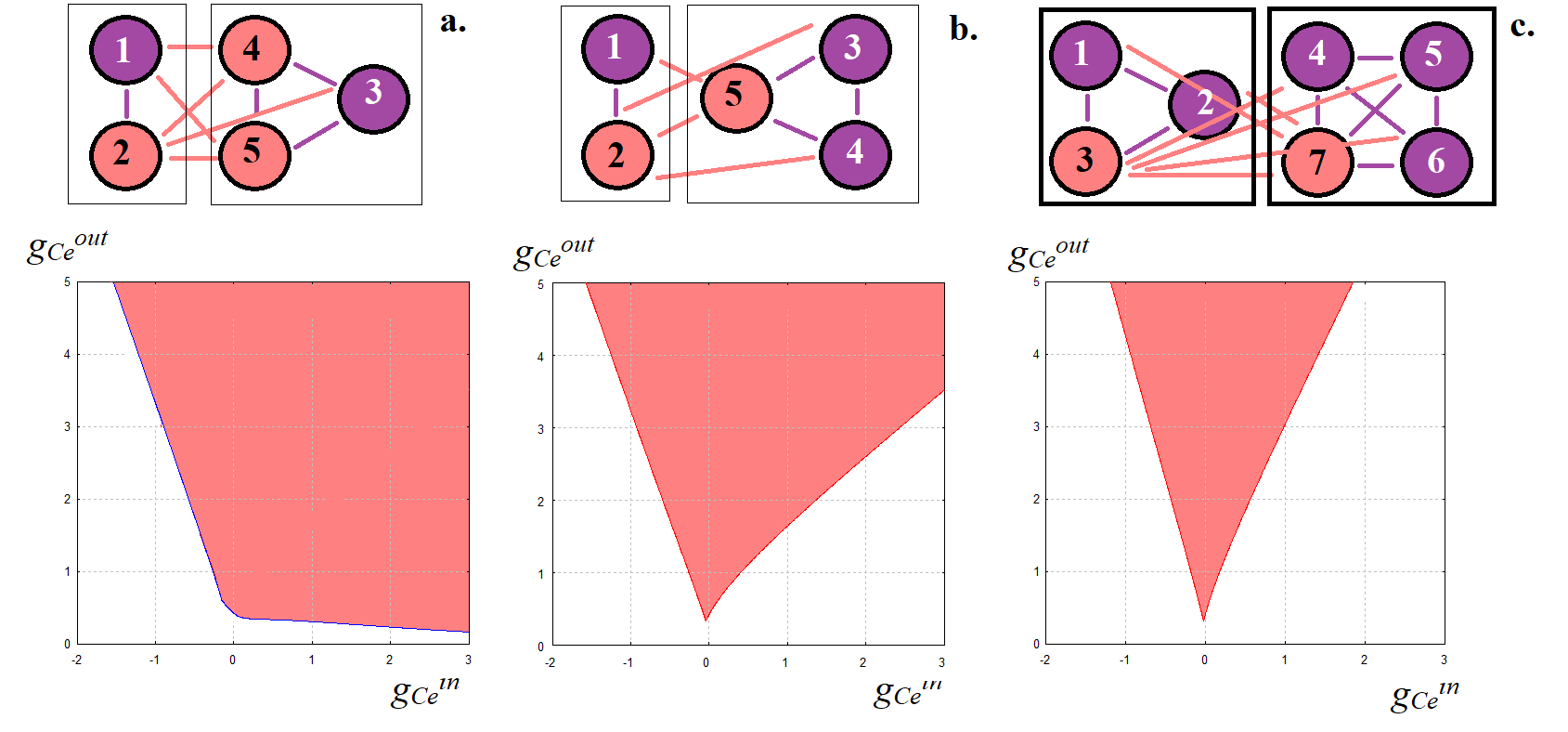}
\caption{Schematic representation of coupled networks and bifurcation diagram for model (\ref{eq:6cell_mn}). Parameters: $\tau$ = 0.02, $\tau_S$ = 35, $\sigma$ = 0.93, $g_{Ca} = 3.6$, $g_K = 10.0$, $g_{K2} = 0.2$, $g_S$ = 4.0, $V_Ca$ = 25.0, $V_K$ = -75.0, $V_m$ = -20.0, $\theta_m$ = 12.0, $V_n$ = -16.0, $\theta_n$ = 5.6, $V_S$ = -35, $\theta_S$ = 10.0, $V_P$ = -47.0, $\theta_p$ = 1.0. $k_1=0$, $k_2=1$, $k_3=0$, $k_4=1$. \textbf{a.} $k_1=0$, $k_2=1$, $k_3=0$, $k_4=1$. \textbf{b.}Red notes have communication defect and demonstrate bistability between silent state and bursting oscillations.}
\label{fig:5}
\end{figure*}

Now let us consider the cases when the number of elements in the subsystems is different. The simplest such situation can be implemented in the interaction of subsystems from two and three models. In this case, the nodes in the first subsystem will be different: one with a communication defect, the other without. And in the second subsystem, we can consider two cases: (i) 1 node with a defect and 2 without (Fig.~\ref{fig:5}a), and vice versa (ii) 2 nodes with a defect and 1 without (Fig.~\ref{fig:5}b). In the second situation, the number of defective elements will exceed normal ones, and stabilization in the global network is also possible. Figure \ref{fig:5} shows the results of a numerical bifurcation analysis of such a system. Equilibrium state stabilization is possible in such a network, as well as for subnets of the same size. The area on the plane of communication parameters in the case when there are fewer elements without pathology than elements with pathology are less. The equilibrium stabilization threshold in a globally coupled network is 2.334. Taking into account the multiplexity of the network, this threshold becomes even higher.

Figure \ref{fig:5}c shows illustrations for a larger network ($N = 7$), but with a minimum number of nodes in the network (under one in the subnets and, accordingly, 2 in the whole network). In such a network, it is also possible to stabilize a stable equilibrium state. The stabilization threshold increases with an increase in the size of the network, the area in the space of parameters decreases.

\section{\label{sec:discus}Conclusions}

The paper presents a study of the dynamics of a network of oscillators. Two types of very similar models based on the Hodgkin-Huxley formalism are used as the basic elements of the network. The first type model demonstrates bursting oscillations. The second model, like the first, demonstrates bursting oscillations, but in the phase space of a system with a bursting attractor, a stable equilibrium state coexists; this feature is caused by taking into account an additional ion channel, which has a non-monotonic characteristic and can be interpreted as a channel with a communication defect. The work investigated heterogeneous networks, some of the nodes of which had a communication defect, and some did not. In previous works, it was shown that heterogeneous networks of this type are quite resistant to the influence of nodes with a communication defect. Stabilization of the equilibrium state is possible only if the number of network elements with a defect dominates. In this paper, multiplex networks were developed and investigated, assuming more active communication between models with a defect. The work shows that with such a connection, it is enough to have one element with a communication defect in the subnetworks in order to stabilize the state of equilibrium in the entire network.

\section*{Acknowledgements}

This work is supported by a grant from the Russian Science Foundation (No. 20-71-10048). This research was leaded in collaboration with A. Koseska, author thanks for her interest to the work.

\section*{Data Availability}
The data supporting numerical experiments presented in this paper are available from the corresponding author upon reasonable request.

\nocite{*}

\begin{thebibliography}{16}%
\makeatletter
\providecommand \@ifxundefined [1]{%
 \@ifx{#1\undefined}
}%
\providecommand \@ifnum [1]{%
 \ifnum #1\expandafter \@firstoftwo
 \else \expandafter \@secondoftwo
 \fi
}%
\providecommand \@ifx [1]{%
 \ifx #1\expandafter \@firstoftwo
 \else \expandafter \@secondoftwo
 \fi
}%
\providecommand \natexlab [1]{#1}%
\providecommand \enquote  [1]{``#1''}%
\providecommand \bibnamefont  [1]{#1}%
\providecommand \bibfnamefont [1]{#1}%
\providecommand \citenamefont [1]{#1}%
\providecommand \href@noop [0]{\@secondoftwo}%
\providecommand \href [0]{\begingroup \@sanitize@url \@href}%
\providecommand \@href[1]{\@@startlink{#1}\@@href}%
\providecommand \@@href[1]{\endgroup#1\@@endlink}%
\providecommand \@sanitize@url [0]{\catcode `\\12\catcode `\$12\catcode
  `\&12\catcode `\#12\catcode `\^12\catcode `\_12\catcode `\%12\relax}%
\providecommand \@@startlink[1]{}%
\providecommand \@@endlink[0]{}%
\providecommand \url  [0]{\begingroup\@sanitize@url \@url }%
\providecommand \@url [1]{\endgroup\@href {#1}{\urlprefix }}%
\providecommand \urlprefix  [0]{URL }%
\providecommand \Eprint [0]{\href }%
\providecommand \doibase [0]{http://dx.doi.org/}%
\providecommand \selectlanguage [0]{\@gobble}%
\providecommand \bibinfo  [0]{\@secondoftwo}%
\providecommand \bibfield  [0]{\@secondoftwo}%
\providecommand \translation [1]{[#1]}%
\providecommand \BibitemOpen [0]{}%
\providecommand \bibitemStop [0]{}%
\providecommand \bibitemNoStop [0]{.\EOS\space}%
\providecommand \EOS [0]{\spacefactor3000\relax}%
\providecommand \BibitemShut  [1]{\csname bibitem#1\endcsname}%
\let\auto@bib@innerbib\@empty
\bibitem [{\citenamefont {Bornholdt}\ and\ \citenamefont
  {Schuster}(2001)}]{bornholdt2001handbook}%
  \BibitemOpen
  \bibfield  {author} {\bibinfo {author} {\bibfnamefont {S.}~\bibnamefont
  {Bornholdt}}\ and\ \bibinfo {author} {\bibfnamefont {H.~G.}\ \bibnamefont
  {Schuster}},\ }\bibfield  {title} {\enquote {\bibinfo {title} {Handbook of
  graphs and networks},}\ }\href@noop {} {\bibfield  {journal} {\bibinfo
  {journal} {From Genome to the Internet, Willey-VCH (2003 Weinheim)}\ }
  (\bibinfo {year} {2001})}\BibitemShut {NoStop}%
\bibitem [{\citenamefont {Strogatz}(2001)}]{strogatz2001exploring}%
  \BibitemOpen
  \bibfield  {author} {\bibinfo {author} {\bibfnamefont {S.~H.}\ \bibnamefont
  {Strogatz}},\ }\bibfield  {title} {\enquote {\bibinfo {title} {Exploring
  complex networks},}\ }\href@noop {} {\bibfield  {journal} {\bibinfo
  {journal} {nature}\ }\textbf {\bibinfo {volume} {410}},\ \bibinfo {pages}
  {268--276} (\bibinfo {year} {2001})}\BibitemShut {NoStop}%
\bibitem [{\citenamefont {Dorogovtsev}\ and\ \citenamefont
  {Mendes}(2002)}]{dorogovtsev2002evolution}%
  \BibitemOpen
  \bibfield  {author} {\bibinfo {author} {\bibfnamefont {S.~N.}\ \bibnamefont
  {Dorogovtsev}}\ and\ \bibinfo {author} {\bibfnamefont {J.~F.}\ \bibnamefont
  {Mendes}},\ }\bibfield  {title} {\enquote {\bibinfo {title} {Evolution of
  networks},}\ }\href@noop {} {\bibfield  {journal} {\bibinfo  {journal}
  {Advances in physics}\ }\textbf {\bibinfo {volume} {51}},\ \bibinfo {pages}
  {1079--1187} (\bibinfo {year} {2002})}\BibitemShut {NoStop}%
\bibitem [{\citenamefont {Newman}(2003)}]{newman2003structure}%
  \BibitemOpen
  \bibfield  {author} {\bibinfo {author} {\bibfnamefont {M.~E.}\ \bibnamefont
  {Newman}},\ }\bibfield  {title} {\enquote {\bibinfo {title} {The structure
  and function of complex networks},}\ }\href@noop {} {\bibfield  {journal}
  {\bibinfo  {journal} {SIAM review}\ }\textbf {\bibinfo {volume} {45}},\
  \bibinfo {pages} {167--256} (\bibinfo {year} {2003})}\BibitemShut {NoStop}%
\bibitem [{\citenamefont {Kirst}, \citenamefont {Timme},\ and\ \citenamefont
  {Battaglia}(2016)}]{kirst2016dynamic}%
  \BibitemOpen
  \bibfield  {author} {\bibinfo {author} {\bibfnamefont {C.}~\bibnamefont
  {Kirst}}, \bibinfo {author} {\bibfnamefont {M.}~\bibnamefont {Timme}}, \ and\
  \bibinfo {author} {\bibfnamefont {D.}~\bibnamefont {Battaglia}},\ }\bibfield
  {title} {\enquote {\bibinfo {title} {Dynamic information routing in complex
  networks},}\ }\href@noop {} {\bibfield  {journal} {\bibinfo  {journal}
  {Nature communications}\ }\textbf {\bibinfo {volume} {7}},\ \bibinfo {pages}
  {1--9} (\bibinfo {year} {2016})}\BibitemShut {NoStop}%
\bibitem [{\citenamefont {Zhang}\ \emph {et~al.}(2020)\citenamefont {Zhang},
  \citenamefont {Witthaut}, \citenamefont {Timme} \emph
  {et~al.}}]{zhang2020topological}%
  \BibitemOpen
  \bibfield  {author} {\bibinfo {author} {\bibfnamefont {X.}~\bibnamefont
  {Zhang}}, \bibinfo {author} {\bibfnamefont {D.}~\bibnamefont {Witthaut}},
  \bibinfo {author} {\bibfnamefont {M.}~\bibnamefont {Timme}},  \emph
  {et~al.},\ }\bibfield  {title} {\enquote {\bibinfo {title} {Topological
  determinants of perturbation spreading in networks},}\ }\href@noop {}
  {\bibfield  {journal} {\bibinfo  {journal} {Physical Review Letters}\
  }\textbf {\bibinfo {volume} {125}},\ \bibinfo {pages} {218301} (\bibinfo
  {year} {2020})}\BibitemShut {NoStop}%
\bibitem [{\citenamefont {Hodgkin}\ and\ \citenamefont
  {Huxley}(1952)}]{hodgkin1952currents}%
  \BibitemOpen
  \bibfield  {author} {\bibinfo {author} {\bibfnamefont {A.~L.}\ \bibnamefont
  {Hodgkin}}\ and\ \bibinfo {author} {\bibfnamefont {A.~F.}\ \bibnamefont
  {Huxley}},\ }\bibfield  {title} {\enquote {\bibinfo {title} {Currents carried
  by sodium and potassium ions through the membrane of the giant axon of
  loligo},}\ }\href@noop {} {\bibfield  {journal} {\bibinfo  {journal} {The
  Journal of physiology}\ }\textbf {\bibinfo {volume} {116}},\ \bibinfo {pages}
  {449--472} (\bibinfo {year} {1952})}\BibitemShut {NoStop}%
\bibitem [{\citenamefont {Izhikevich}(2007)}]{izhikevich2007dynamical}%
  \BibitemOpen
  \bibfield  {author} {\bibinfo {author} {\bibfnamefont {E.~M.}\ \bibnamefont
  {Izhikevich}},\ }\href@noop {} {\emph {\bibinfo {title} {Dynamical systems in
  neuroscience}}}\ (\bibinfo  {publisher} {MIT press},\ \bibinfo {year}
  {2007})\BibitemShut {NoStop}%
\bibitem [{\citenamefont {Shen}\ \emph {et~al.}(2021)\citenamefont {Shen},
  \citenamefont {Zhu}, \citenamefont {Liu},\ and\ \citenamefont
  {Wen}}]{shen2021multistability}%
  \BibitemOpen
  \bibfield  {author} {\bibinfo {author} {\bibfnamefont {Y.}~\bibnamefont
  {Shen}}, \bibinfo {author} {\bibfnamefont {S.}~\bibnamefont {Zhu}}, \bibinfo
  {author} {\bibfnamefont {X.}~\bibnamefont {Liu}}, \ and\ \bibinfo {author}
  {\bibfnamefont {S.}~\bibnamefont {Wen}},\ }\bibfield  {title} {\enquote
  {\bibinfo {title} {Multistability and associative memory of neural networks
  with morita-like activation functions},}\ }\href@noop {} {\bibfield
  {journal} {\bibinfo  {journal} {Neural Networks}\ } (\bibinfo {year}
  {2021})}\BibitemShut {NoStop}%
\bibitem [{\citenamefont {Malashchenko}, \citenamefont {Shilnikov},\ and\
  \citenamefont {Cymbalyuk}(2011)}]{malashchenko2011six}%
  \BibitemOpen
  \bibfield  {author} {\bibinfo {author} {\bibfnamefont {T.}~\bibnamefont
  {Malashchenko}}, \bibinfo {author} {\bibfnamefont {A.}~\bibnamefont
  {Shilnikov}}, \ and\ \bibinfo {author} {\bibfnamefont {G.}~\bibnamefont
  {Cymbalyuk}},\ }\bibfield  {title} {\enquote {\bibinfo {title} {Six types of
  multistability in a neuronal model based on slow calcium current},}\
  }\href@noop {} {\bibfield  {journal} {\bibinfo  {journal} {PLoS One}\
  }\textbf {\bibinfo {volume} {6}},\ \bibinfo {pages} {e21782} (\bibinfo {year}
  {2011})}\BibitemShut {NoStop}%
\bibitem [{\citenamefont {Stankevich}, \citenamefont {Mosekilde},\ and\
  \citenamefont {Koseska}(2018)}]{stankevich2018stochastic}%
  \BibitemOpen
  \bibfield  {author} {\bibinfo {author} {\bibfnamefont {N.}~\bibnamefont
  {Stankevich}}, \bibinfo {author} {\bibfnamefont {E.}~\bibnamefont
  {Mosekilde}}, \ and\ \bibinfo {author} {\bibfnamefont {A.}~\bibnamefont
  {Koseska}},\ }\bibfield  {title} {\enquote {\bibinfo {title} {Stochastic
  switching in systems with rare and hidden attractors},}\ }\href@noop {}
  {\bibfield  {journal} {\bibinfo  {journal} {The European Physical Journal
  Special Topics}\ }\textbf {\bibinfo {volume} {227}},\ \bibinfo {pages}
  {747--756} (\bibinfo {year} {2018})}\BibitemShut {NoStop}%
\bibitem [{\citenamefont {Stankevich}\ and\ \citenamefont
  {Mosekilde}(2017)}]{stankevich2017coexistence}%
  \BibitemOpen
  \bibfield  {author} {\bibinfo {author} {\bibfnamefont {N.}~\bibnamefont
  {Stankevich}}\ and\ \bibinfo {author} {\bibfnamefont {E.}~\bibnamefont
  {Mosekilde}},\ }\bibfield  {title} {\enquote {\bibinfo {title} {Coexistence
  between silent and bursting states in a biophysical hodgkin-huxley-type of
  model},}\ }\href@noop {} {\bibfield  {journal} {\bibinfo  {journal} {Chaos:
  An Interdisciplinary Journal of Nonlinear Science}\ }\textbf {\bibinfo
  {volume} {27}},\ \bibinfo {pages} {123101} (\bibinfo {year}
  {2017})}\BibitemShut {NoStop}%
\bibitem [{\citenamefont {Sherman}, \citenamefont {Rinzel},\ and\ \citenamefont
  {Keizer}(1988)}]{sherman1988emergence}%
  \BibitemOpen
  \bibfield  {author} {\bibinfo {author} {\bibfnamefont {A.}~\bibnamefont
  {Sherman}}, \bibinfo {author} {\bibfnamefont {J.}~\bibnamefont {Rinzel}}, \
  and\ \bibinfo {author} {\bibfnamefont {J.}~\bibnamefont {Keizer}},\
  }\bibfield  {title} {\enquote {\bibinfo {title} {Emergence of organized
  bursting in clusters of pancreatic beta-cells by channel sharing},}\
  }\href@noop {} {\bibfield  {journal} {\bibinfo  {journal} {Biophysical
  journal}\ }\textbf {\bibinfo {volume} {54}},\ \bibinfo {pages} {411--425}
  (\bibinfo {year} {1988})}\BibitemShut {NoStop}%
\bibitem [{\citenamefont {Sherman}\ and\ \citenamefont
  {Rinzel}(1992)}]{sherman1992rhythmogenic}%
  \BibitemOpen
  \bibfield  {author} {\bibinfo {author} {\bibfnamefont {A.}~\bibnamefont
  {Sherman}}\ and\ \bibinfo {author} {\bibfnamefont {J.}~\bibnamefont
  {Rinzel}},\ }\bibfield  {title} {\enquote {\bibinfo {title} {Rhythmogenic
  effects of weak electrotonic coupling in neuronal models},}\ }\href@noop {}
  {\bibfield  {journal} {\bibinfo  {journal} {Proceedings of the National
  Academy of Sciences}\ }\textbf {\bibinfo {volume} {89}},\ \bibinfo {pages}
  {2471--2474} (\bibinfo {year} {1992})}\BibitemShut {NoStop}%
\bibitem [{\citenamefont {Stankevich}\ and\ \citenamefont
  {Koseska}(2020)}]{stankevich2020cooperative}%
  \BibitemOpen
  \bibfield  {author} {\bibinfo {author} {\bibfnamefont {N.}~\bibnamefont
  {Stankevich}}\ and\ \bibinfo {author} {\bibfnamefont {A.}~\bibnamefont
  {Koseska}},\ }\bibfield  {title} {\enquote {\bibinfo {title} {Cooperative
  maintenance of cellular identity in systems with intercellular communication
  defects},}\ }\href@noop {} {\bibfield  {journal} {\bibinfo  {journal} {Chaos:
  An Interdisciplinary Journal of Nonlinear Science}\ }\textbf {\bibinfo
  {volume} {30}},\ \bibinfo {pages} {013144} (\bibinfo {year}
  {2020})}\BibitemShut {NoStop}%
\bibitem [{\citenamefont {Ermentrout}(2002)}]{ermentrout2002simulating}%
  \BibitemOpen
  \bibfield  {author} {\bibinfo {author} {\bibfnamefont {B.}~\bibnamefont
  {Ermentrout}},\ }\href@noop {} {\emph {\bibinfo {title} {Simulating,
  analyzing, and animating dynamical systems: a guide to XPPAUT for researchers
  and students}}}\ (\bibinfo  {publisher} {SIAM},\ \bibinfo {year}
  {2002})\BibitemShut {NoStop}%
\end{thebibliography}
%

\end{document}